\documentclass[prd,showpacs,twocolumn,superscriptaddress,floatfix,nofootinbib,10pt]{revtex4-2}
\usepackage{setspace}
\usepackage[utf8]{inputenc}
\usepackage{float}
\usepackage{amsmath,amssymb,amsfonts,bm}
\usepackage{graphicx}
\usepackage{multirow}
\usepackage[usenames,dvipsnames]{color}
\allowdisplaybreaks
\usepackage[
colorlinks=true,
linkcolor=blue,
breaklinks=true,
urlcolor=blue,
citecolor=blue]{hyperref}

\usepackage{orcidlink}
\usepackage{subfigure}
\usepackage{soul}
\usepackage{xcolor}
\usepackage{physics}
\usepackage{booktabs}
\usepackage{ulem}
\usepackage{tikz-feynman}
\usepackage{makecell}
\definecolor{green}{rgb}{0,0.6,0}
\newcommand{\mev}{\textrm{ MeV}}

\newcommand{\GXNU}{\affiliation{Department of Physics, Guangxi Normal University, Guilin 541004, China}}

\newcommand{\GXZD}{\affiliation{Guangxi Key Laboratory of Nuclear Physics and Technology, Guangxi Normal University, Guilin 541004, China}}

\newcommand{\IFIC}{\affiliation{Departamento de F\'{\i}sica Te\'orica and IFIC, Centro Mixto Universidad de
		Valencia-CSIC Institutos de Investigaci\'on de Paterna, Apartado 22085,
		46071 Valencia, Spain}}

\begin{document}
	\title{Superexotic $K^{*+}D^{*+}K^{*+}$ bound state}
	
	\begin{abstract}
	We study a system made from $K^{*+}D^{*+}K^{*+}$ with charge $3$, isospin $I=3/2$, spin $J=3$, and a quark content of $c\bar d \bar s u \bar s u$, which make it highly exotic relative to the standard $q\bar q$ structure of mesons. The interaction of the three body system is obtained starting from a cluster of $D^{*+} K^{*+}$ in $I=1$ and $J=2$, that in different works has been found bound, and adding to it an extra $K^{*+}$ with spin aligned with those of the vectors of the cluster. We find that the $K^* K^*$ interaction in $I=1$ and $J=2$ is repulsive, but its strength is small compared to that of $D^{*+} K^{*+}$ in $I=1$ and $J=2$, such that we find a three body state bound by about $100 \, \rm MeV$ with respect to the mass of a $K^{*+}$ and the $D^{*+} K^{*+}$ cluster. The width of the state, of about $10\, \rm MeV$, is much smaller than the binding, which facilitates its observation. We suggest to find that state by measuring the invariant mass of $K D K^*$, something feasible in present experimental facilities.
	\end{abstract}
	
	\author{Wen-Hao Jia}
	\GXNU%

	\author{Pei-Shen Su}
	\GXNU%
	
	\author{Wei-Hong Liang\orcidlink{0000-0001-5847-2498}}%
	\email{liangwh@gxnu.edu.cn}
	\GXNU%
	\GXZD%

	\author{Raquel Molina}%
	\email{Raquel.Molina@ific.uv.es}
	\IFIC%

	\author{Eulogio Oset}%
	\email{Oset@ific.uv.es}
	\GXNU%
	\IFIC%

\maketitle

\section{Introduction}\label{sec:Intr}
Exotic meson states which do not follow the standard $q\bar{q}$ nature have been reported during the last years and are the subject of intense attention~\cite{Esposito:2016noz,Guo:2017jvc,Olsen:2017bmm,Lebed:2016hpi,Chen:2016qju,Liu:2019zoy,Brambilla:2019esw,Ali:2017jda,Karliner:2017qhf,Guo:2013sya,Wu:2022ftm,Meng:2022ozq,Liu:2024uxn}.
Many of these exotic states qualify as molecular states of pairs of elementary mesons~\cite{Guo:2017jvc}.
The possibility of having bound states of three mesons has also been exploited, but due to the extra difficulties of experimental observation and more involved theoretical calculations, the field has received far less attention than the molecular states of two particles.
Even then, 30 cases of such systems have been reported in the review of Ref.~\cite{MartinezTorres:2020hus}.
Work has continued reporting on the $D^*D^*\bar{K}^*$ molecule~\cite{Ikeno:2022jbb}, a $D\bar{D}K$ bound state in Ref.~\cite{Wu:2022wgn}, the $\eta K^*\bar{K}^*$, $\pi K^*\bar{K}^*$, and $K K^*\bar{K}^*$ systems in Ref.~\cite{Shen:2022etd}, the state of triple charm $D^*D^*D$ in Ref.~\cite{Wu:2021kbu,Luo:2021ggs}, the $D^*D^*D^*$ in Ref.~\cite{Luo:2021ggs,Bayar:2022bnc}, the $K^*B^*B^*$ in Ref.~\cite{Bayar:2023itf}, the $DD^*K$ in Ref.~\cite{Zhang:2024yfj}, the $KD^{(*)}D^*$ in Ref.~\cite{Ren:2024mjh}, the $DDD^*$ in Ref.~\cite{Zhu:2024hgm} and the $\bar{D}_sDK$ in Ref.~\cite{Wu:2025fzx}.
The $D^*D^*D^*$ system has also been discussed concerning the possibility of finding Efimov states in three-body hadronic systems~\cite{Ortega:2024ecy,Fu:2025joa}.

In the present work, we propose to look for a superexotic state that we find bound: the $K^{*+}D^{*+}K^{*+}$ system with total spin $J = 3$.
The system has the quark content $c \bar{d} u \bar{s} u \bar{s}$, and flavor  conservation in strong interactions prevents it to decay into any two mesons, adding stability to the system.
It has isospin $\frac{3}{2} $ and charge $3$, which no $q\bar{q}$ meson has, and contains necessarily $6$ quarks, since no annihilation of $q\bar{q}$ with the same flavor is possible. In addition, it has $J = 3$, with no orbital excitation of the quarks involved in the system.

Multimeson molecular states of high spin have been reported before, and in Ref.~\cite{Roca:2010tf} the physical states $\rho_3(1690)$, $f_4(2050)$, $\rho_5(2350)$, and $f_6(2510)$ were generated as multirho states with spins aligned.
Similarly, in Ref.~\cite{Yamagata-Sekihara:2010muv}, the $K^*_3(1780)$, $K_4^*(2045)$, and $K_5^*(2380)$ were generated as $K$-multirho states, and in Ref.~\cite{Xiao:2012dw}, several states with a $D$ and multirho with high spins were also predicted.

The reason to propose the $K^{*+}D^{*+}K^{*+}$ system stems from the fact that the $K^*D^*$ interaction with isospin $I = 1$ and spin $J = 2$ is predicted to be attractive and leads to a bound state in Ref.~\cite{Molina:2010tx}, using as input the $K^*D^*$ and $D^*_s \rho$ coupled channels with their interaction given by the local hidden gauge approach~\cite{Bando:1984ej,Bando:1987br,Meissner:1987ge,Nagahiro:2008cv}.
In the $I = 1$, $J = 0,\,1$ cases, the interaction was found weaker, and only a cusp in the amplitudes was observed.
The experimental observation of the $X_0(2866)$ state (now named $T_{c\bar s}(2900)$)~\cite{LHCb:2020bls,LHCb:2020pxc} brought new light to this problem.
Indeed, that state, with $I = 1$, $C = 1$, $S = 1$, and $J = 0$, could be associated with the cusp found in Ref.~\cite{Molina:2010tx}, and in Ref.~\cite{Molina:2022jcd} the problem was revisited, introducing new decay channels, in particular the $D_s^+ \pi^+$ where it was observed, and fine-tuning the input of the theory to match the experimental $D_s^+ \pi^+$ mass distribution in the $B \to \bar{D} D_s \pi$ decay.
The experimental information is found consistent with the findings of Ref.~\cite{Molina:2010tx}, and once again, the $T_{c\bar s}(2900)$ is associated in Ref.~\cite{Molina:2022jcd} to a virtual state created by the $D_s^* \rho$, $D^* K^*$ interaction in coupled channels, which materializes as a strong cusp in the amplitude.
The interesting issue is that both in Ref.~\cite{Molina:2010tx} and Ref.~\cite{Molina:2022jcd}, the $J = 2$ state with the same other quantum numbers is found bound, and in Ref.~\cite{Molina:2022jcd}, a state with a mass of $2834\,\mathrm {MeV}$ (a binding energy of $48\,\mathrm {MeV}$ with respect to $D_s^* \rho$ and $68\,\mathrm {MeV}$ with respect to $D^* K^*$) and $\Gamma = 19\,\mathrm {MeV}$ was predicted.

The nature of $T_{cs0}$ as a $D^*K^*$ molecule has also been supported by other independent work~\cite{Yue:2022mnf,Agaev:2022eyk,Ortega:2023azl,Duan:2023lcj,Wang:2023hpp}, although other tetraquark configurations of non-molecular nature have also been proposed~\cite{Liu:2022hbk,Yang:2023wgu}.
In Ref.~\cite{Duan:2023lcj}, the interaction in the $J = 2$ channel is also found more attractive, and a bound state is also found.

It is interesting to compare the situation here with that of the $DK$ interaction in $I=1$. 
Using QCD lattice data it was inferred in Ref. \cite{Liu:2012zya} that $DK$ in $I=0$ was very attractive, leading to the formation of a bound state identified with the $D_{s0}^*(2317)$ resonance, however this was not the case for $DK$ in $I=1$. 
This is in spite of having the accompanying $D_s \pi$ channel which can help producing an attraction \cite{Hyodo:2013nka,Aceti:2014ala,Wang:2022pin}. However, the $D_s \pi$ channel is far away from $DK$ and this weakens the possible attraction.  
Actually, this is also the case when using effective theory with the interaction induced by the local hidden gauge approach \cite{Ikeno:2023ojl}, where one sees that the $DK$ interaction is attractive in $I=0$ but repulsive in $I=1$. 
Further steps including the $D_s \pi$ channel done in Ref.~\cite{Su:2025aiz} lead to the strong decay of the $D_{s0}^*(2317)$ but do not produce enough attraction to generate a bound state in $I=1$. 

The situation is different in the case of the $D^* K^*$ interaction in $I=1$ because the $D^*_s \rho$ coupled channel is much closer to the $D^* K^*$ threshold and produces an attraction that leads to the mentioned strong cusp in spin $J=0$. Then in $J=2$ the attraction found is stronger and leads to the state predicted in Refs.~\cite{Molina:2010tx,Molina:2022jcd,Duan:2023lcj}.

With this solid backing for the strong attraction of $D^*K^*$ in $I=1$ and $J=2$, we can tackle the study of the $K^*D^*K^*$ system with $J=3$.
By putting all the spins of the vectors aligned, we shall have $J=3$ and all pairs will have $J=2$. Then, we benefit from twice the $D^*K^*$ interaction in the $J=2$ channel.
The $K^*K^*$ interaction in $I=1$, $J=2$ will turn out to be repulsive in the calculation that we shall perform here, but we shall see that its strength is not enough to overcome the strong attraction of the other two pairs, and as a consequence, we obtain a clean bound state of the three-body system.

\section{formalism}\label{sec:form}
We will follow the fixed center approximation (FCA)~\cite{Foldy:1945zz,Brueckner:1953zz,Brueckner:1953zza,Chand:1962ec,Barrett:1999cw,Deloff:1999gc,Kamalov:2000iy,MartinezTorres:2020hus,Roca:2010tf,Malabarba:2024hlv}, to study the interaction of the three-body system. 
For that we choose a cluster of two particles, which already gives a bound state. 
This cluster is the $K^{*+}D^{*+}$ system with $J=2$, which is bound by $68 \mev$ with respect to the $K^*D^*$ threshold, as was mentioned in the introduction. 
Then, the third particle, the other $K^{*+}$, will interact with the cluster. 
We follow recent developments where the FCA method has been improved to satisfy elastic unitarity at threshold of the third particle and the cluster~\cite{Ikeno:2025bsx,Agatao:2025ckp}.
This condition is necessary to evaluate correlation functions for the interaction of a particle with a bound system of two other particles, as discussed in Ref.~\cite{Encarnacion:2025lyf}, and also has repercussion in the binding of the three-body system.

In Ref.~\cite{Ikeno:2025bsx}, the interaction of a neutron $n$ with a $KD$ cluster, which forms the $D_{s0}^*(2317)$, was studied. 
Since the isospin of the particles is the same as in the present case, and the $nK,\,nD$ interaction in $S$-wave is spin independent, we can use the results of Ref.~\cite{Ikeno:2025bsx,Agatao:2025ckp} with minor modifications.
The only needed change is due to the fact that we follow the normalization of Mandl and Shaw~\cite{mandl_quantum_2010} for the fields and propagators. 
Thus, we must change
\begin{equation}
	\begin{split}
	\frac{M}{E(\vec{q}\,)}\,\frac{1}{q^0 - E(\vec{q}\,) + i\epsilon}
	\;\; \longrightarrow \;\;&
	\frac{1}{q_0^2 - \vec{q}^{\,2} - m^2}
	\\
	\simeq &
	\frac{1}{2\omega(\vec{q}\,)}\,
	\frac{1}{q^0 - \omega(\vec{q}\,) + i\epsilon},
	\end{split}
	\label{eq:1}
\end{equation}
where $M$ is the mass of the neutron in Ref.~\cite{Ikeno:2025bsx}, $m$ the mass of the $K^*$ here, and $E(\vec{q}\,) = \sqrt{M^2 + \vec{q}^{\,2}}$, $\omega(\vec{q}\,) = \sqrt{m^2 + \vec{q}^{\,2}}$.

We refrain from repeating the formalism of Ref.~\cite{Ikeno:2025bsx}, and we directly use the results from there with the modifications of Eq.~\eqref{eq:1}. 
The FCA formalism sums the diagrams of Fig.~1 of Ref.~\cite{Ikeno:2025bsx}, substituting $n$ by $K^*$ and $D, K$ by $D^*, K^*$, respectively. 
In addition, the unitary amplitude sums the diagrams of Fig.~2 of Ref.~\cite{Ikeno:2025bsx}. 
The sum of all those diagrams is obtained in Eq.~(29) of Ref.~\cite{Ikeno:2025bsx}, which is rewritten in a simplified form in Ref.~\cite{Agatao:2025ckp}, and reads
\begin{equation}
	T=\frac{\tilde{t}_{1}+\tilde{t}_{2}+\left(2 G_{0}-G_{C}^{(1)}-G_{C}^{(2)}\right) \tilde{t}_{1}\, \tilde{t}_{2}}{1-G_{C}^{(1)} \,\tilde{t}_{1}-G_{C}^{(2)} \,\tilde{t}_{2}-\left(G_{0}^{2}-G_{C}^{(1)} \, G_{C}^{(2)}\right) \tilde{t}_{1} \,\tilde{t}_{2}},
	\label{eq:2}
\end{equation}
where, for reasons of normalization, $\tilde{t}_1$, $\tilde{t}_2$ are defined as
\begin{equation}
	\tilde{t}_1 = \frac{M_c}{M_{D^*}} \,t_1 , \quad
	\tilde{t}_2 = \frac{M_c}{M_{K^*}} \,t_2,
	\label{eq:3}
\end{equation}
where $t_1$, $t_2$ are the scattering amplitudes of the external $K^{*+}$ with the $D^{*+}$ and $K^{*+}$ of the cluster, respectively, in $I = 1$ and $J = 2$, which are described in the Appendix here, and $M_c$ is the mass of the cluster.
The functions $G_0$, $G_C^{(1)}$, and $G_C^{(2)}$ stand for the external $K^*$ propagator folded by the cluster wave function, which involve the cluster form factors. $G_0$ represents the $K^*$ propagation from particle $1$ to particle $2$ of the cluster, and $G_C^{(1)}$, $G_C^{(2)}$ for the $K^*$ propagation together with the cluster as a whole, from particle 1 to particle 1 ($G_C^{(1)}$) and from particle 2 to particle 2 ($G_C^{(2)}$).

These $G$ functions are given by
\begin{equation}
	\begin{split}
	G_0(\sqrt{s}) =&
	\int \frac{d^3 q}{(2\pi)^3} \,
	\frac{F_C(\vec q\,)}{\sqrt{s} - \omega_{K^*}(\vec q \,) - \omega_C(\vec q \,) + i\epsilon} \frac{1}{2\omega_{K^*}(\vec q \,)} \,\\
	&\cdot \frac{1}{2\omega_C(\vec q \,)} \,
	\Theta(q_{\text{max}}^{(1)} - q_1^*) \,
	\Theta(q_{\text{max}}^{(2)} - q_2^*),
	\end{split}
	\label{eq:4}
\end{equation}
\begin{equation}
	\begin{split}
	G_C^{(i)}(\sqrt{s}) =&
	\int \frac{d^3 q}{(2\pi)^3} \,
	\frac{[F_C^{(i)}(\vec q\,)]^2}{\sqrt{s} - \omega_{K^*}(\vec q \,) - \omega_C(\vec q\,) + i\epsilon} \\
	&\cdot 	\frac{1}{2\omega_{K^*}(\vec q \,)} \,
	\frac{1}{2\omega_C(\vec q \,)} \,
	\Theta(q_{\text{max}}^{(i)} - q_i^*),
	\end{split}
	\label{eq:5}
\end{equation}
where $q_i^*$ are defined later in Eq.~\eqref{eq:10}, $\omega_{K^*}(\vec q\,) = \sqrt{m_{K^*}^2 + \vec{q}^{\,2}}$, $\omega_C(\vec q\,) = \sqrt{M_C^2 + \vec{q}^{\,2}}$, and $F_C^{(i)}(\vec q\,)$ are given by
\begin{equation}
	\begin{split}
	F_C^{(1)}(\vec{q}\,) = F_C\!\left( \frac{M_{K^*}}{M_{D^*} + M_{K^*}} \, \vec{q} \right),\\
	F_C^{(2)}(\vec{q}\,) = F_C\!\left( \frac{M_{D^*}}{M_{D^*} + M_{K^*}} \, \vec{q} \right),
	\end{split}
	\label{eq:6}
\end{equation}
with $F_C(\vec q\,)$ the ordinary form factor of the cluster given by
\begin{equation}
	\begin{split}
	F_C(\vec q\,) =& \frac{F(\vec q\,)}{N},\\
	F(\vec q\,) =&
	\int \frac{d^3 p}{(2\pi)^3} \;
	\frac{1}{M_C - \omega_{D^*}(\vec{p}\,) - \omega_{K^*}(\vec{p}\,)}
	\\
	&\cdot \frac{1}{M_C - \omega_{D^*}(\vec{p} - \vec{q}\,) - \omega_{K^*}(\vec{p} - \vec{q}\,)},
	\end{split}
	\label{eq:7}
\end{equation}
with integration limits $|\vec{p}\,| < q_{\text{max}}$ and $|\vec{p} - \vec{q}\,| < q_{\text{max}}$,
\begin{equation}
	\begin{split}
	N =& F(0) \\
	=& \int_{|\vec{p}\,| < q_{\text{max}}} \frac{d^3 p}{(2\pi)^3}
	\left( \frac{1}{M_C - \omega_{D^*}(\vec{p}\,) - \omega_{K^*}(\vec{p}\,)} \right)^2,
	\end{split}
	\label{eq:8}
\end{equation}
where $q_{\text{max}}$ is the regulator of the $D^* K^*$ loop functions $G$ in the scattering matrix $T = [1 - V G]^{-1} V$ that gives the pole for the $D^* K^*$ state with $J=2$ in Ref.~\cite{Molina:2022jcd}.
With the ingredients used in the Appendix we need $q_{\text{max}} = 967\mev$ to obtain the binding of Ref.~\cite{Molina:2022jcd}.

In addition to the replacement of Eq.~\eqref{eq:1} in the $G_0$ and $G_C^{(i)}$ functions of Eq.~\eqref{eq:4} and Eq.~\eqref{eq:5}, we have implemented the factor $\Theta\bigl(q_{\text{max}}^{(i)} - q^{*}_i\bigr)$ that we justify below.

In the chiral unitary approach to obtain the meson-meson scattering amplitudes~\cite{Oller:1997ti,Kaiser:1998fi,Markushin:2000fa,Nieves:1998hp}, it is common to regularize the loop $G$ functions for two particle propagation with a cut-off $q_{\text{max}}$. The framework is shown equivalent to using a separable potential
\begin{equation}
	V(\vec{q}, \vec{q}\,') = v\,\Theta(q_{\text{max}} - |\vec{q}\,|)\,\Theta(q_{\text{max}} - |\vec{q}\,'|),
	\label{eq:9}
\end{equation}
where $q_{\text{max}}$ gives an idea of the range of the interaction in momentum space, of the order of magnitude of vector meson masses~\cite{Gamermann:2009uq,Song:2022yvz}.
Then, the scattering matrix $T(\vec{q}, \vec{q}\,')$ comes also factorized as $t\,\Theta(q_{\text{max}} - |\vec{q}\,|)\,\Theta(q_{\text{max}} - |\vec{q}\,'|)$.
The formula used for the form factor is demanded by the underlying theory, where one can obtain the formulas for the scattering matrix starting with the potential of Eq.~\eqref{eq:9}. 
This leads to a $t$ matrix containing the two $\Theta$ functions, as we have mentioned, and a wave function in momentum space \cite{Gamermann:2009uq,Yamagata-Sekihara:2010kpd}
\begin{equation*}
	\Psi(p) = g \;\Theta(q_{\rm max} - |\vec p\,|) /[ M_c-\omega_{D^*} (\vec p\,) -\omega_{K^*} (\vec p\,)],
\end{equation*}
with $g$ the coupling of the state to the components of the molecule.
From there the form factor of Eq.~\eqref{eq:7} is readily calculated.

The momenta in the $\Theta$ functions are in the two particle rest frame.
We would like to implement this factor also in our amplitudes $t_1$, $t_2$, for which we must place the momenta of the amplitudes in the rest frame of the external particle and particle 1 of the cluster for $t_1$, and in the rest frame of the external particle and particle 2 of the cluster for $t_2$.
The effect of these cut-offs is not drastic once the natural cluster form factors are used, but we implement them for consistency.
In this case, and given that the momenta involved are relatively small, we find it sufficient to make a Galilean transformation from the frame where the amplitudes appear to the corresponding two-particle rest frame.  
Even then, one has to specify the momenta of the particles of the cluster, and for this we take a prescription used in the study of elastic scattering or coherent production in nuclei, where we start with a momentum for the particle of the cluster $\frac{1}{2} (\vec{q} - \vec{k}\,)$ and $\vec{k}$ for the external particle, and finish with $\frac{1}{2}(\vec{k} - \vec{q}\,)$ for the particle of the cluster and $\vec{q}$ for the external particle.
This symmetrical distribution is the most likely and ensures the smallest momentum components from the cluster wave function~\cite{Carrasco:1991we}.
A different argumentation, reaching the same result, is shown in Ref.~\cite{Boffi:1991nh}.
Then assuming that $\vec{k} \simeq 0$, we find for the rest-frame momenta
\begin{equation}
	\vec{q}^{\,*}_i = \vec{q}\left( 1 - \frac{1}{2} \frac{m_{ex}}{m_{ex} + m'_{i}} \right),
	\label{eq:10}
\end{equation}
with $m_{ex} = m_{K^{*+}}$ and $m'_{1} = m_{D^{*+}}$, $m'_{2} = m_{K^{*+}}$.

\section{Results} \label{sec:res}
In the interaction of two elementary particles, to identify bound states or resonances, it is customary to look for poles in the complex plane. 
Here this task is more cumbersome because in Eq.~\eqref{eq:2}, apart from the $G$ functions which are commonly extrapolated to the unphysical Riemann sheets to search for poles, one has also two-body amplitudes, $t_1$ and $t_2$, which are generally complex. 
This is the reason why in these studies one looks for bound states by looking at the amplitudes in the real axis, which provide all the physical information in any case.

\begin{figure}[t]
\begin{center}
\includegraphics[width=0.48\textwidth]{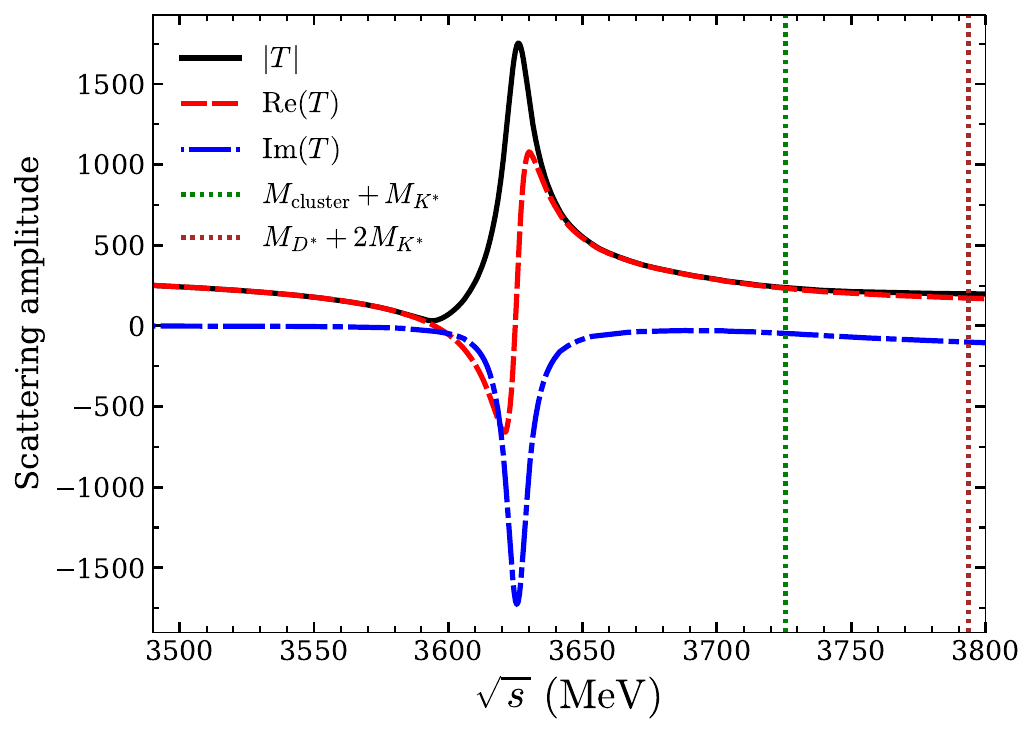}
\end{center}
\vspace{-0.5cm}
\caption{Amplitude $T$ of the three-body $(D^{*+}K^{*+})K^{*+}$ system as a function of the total center-of-mass energy. The black solid, red dashed, and blue dash-dotted lines denote $|T|$, $\mathrm{Re}(T)$, and $\mathrm{Im}(T)$, respectively. The vertical lines indicate the $(D^{*+}K^{*+})\,K^{*+}$ and $D^{*+} K^{*+} K^{*+}$ thresholds.}
\label{Fig:fig1}
\end{figure}

In Fig.~\ref{Fig:fig1} we show the results for the $T$ matrix. 
We separate the real and the imaginary parts, and we see that at $3626 \mev$ we have the typical structure of a resonance, with the real part going through zero when the modulus of the imaginary part has its maximum. 
From the width of $\Im T$ at half strength of its peak, we find that the width of the state obtained is of the order of $10 \mev$, originating from the widths of the $K^*$ and $\rho$ states (mostly from the $\rho$) that have been considered explicitly in the calculation of the two body amplitudes by using the convolution of $G$ function in Eq.~\eqref{eq:A5} of the Appendix.
Note that while the system considered is $K^*D^*K^*$, the input to evaluate the three-body amplitude of this system requires the $K^*D^*$ amplitude discussed in the Appendix, which is calculated using the coupled channels $K^* D^*$ and $D_s^* \rho$. 
Hence the imaginary part of the amplitudes is tied to the widths of the $K^*$ and the $\rho$.
In the conclusion we also mention another similar source of width coming from the decay of the two-body system $K^*D^*$ to $KD$.
We find that the binding of the state with respect to the $K^{*+}$-cluster threshold ($M_{\rm cluster}+M_{K^*}$) is about $100\mev$. 
One can understand this binding from the binding of the external $K^{*+}$ with the $D^{*+}$ of the cluster in $J=2$. 
Indeed, the cluster of $K^{*+} D^{*+}$ with $J=2$ is already bound by $68 \mev$ with respect to the $K^{*+} D^{*+}$ threshold. 
One might rightly ask why we find a larger binding in the particle cluster than in the two particle case. 
The reason is that in the former case the binding is written in terms of the variable $\sqrt{s}$ not $\sqrt{s_1}$ (see Eq.~\eqref{eq:A15} in the Appendix). 
This is explained in Ref.~\cite{Montesinos:2024eoy} and is a feature occurring in the study of the interaction of particles with nuclei, when the binding is written in terms of the energy of the external particle and the nucleus in their rest frame instead of the invariant mass of the external particle with one nucleon of the nucleus. 
\begin{figure}[b]
\begin{center}
\includegraphics[width=0.48\textwidth]{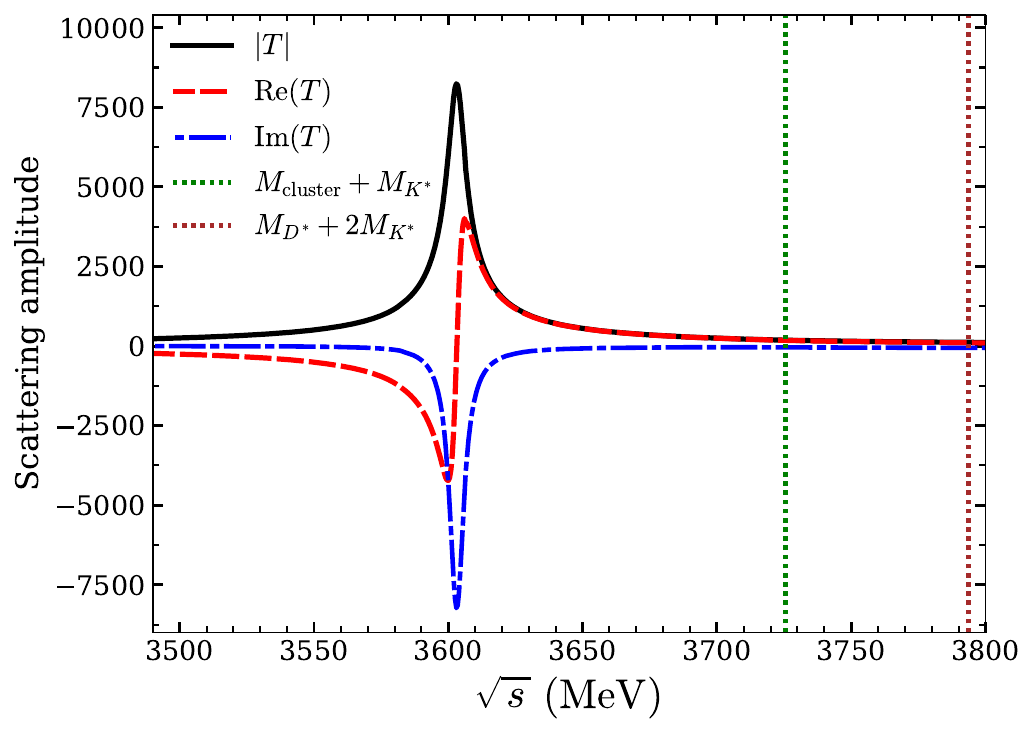}
\end{center}
\vspace{-0.5cm}
\caption{Same as Fig.~\ref{Fig:fig1} but with the $K^{*+}K^{*+}$ interaction switched off ($t_{2}=0$), illustrating the impact of the repulsive $K^{*+}K^{*+}$ force on the three-body amplitude.}
\label{Fig:fig2}
\end{figure}

We should note that the binding energy is in line with other cases of molecular states. 
For instance the $D_{s0}^*(2317)$ is considered basically a $DK$ molecule in effective theories and lattice QCD results, as mentioned above, with a binding of around $45 \mev$. 
Here it is $68 \mev$ for the $D^*K^*$ in $I=1, J=2$, which is of the same order. 
One should note that there are other cases of molecular states where the binding energy is even larger, like the $f_2(1270)$, identified as a $\rho \rho$ bound state in Refs.~\cite{Molina:2008jw,Geng:2008gx}, which has been tested in radiative decays. 
This state is bound by $270 \mev$ considering the nominal mass of the $\rho$, although a large part of the spectrum of the system is less bound when the mass distribution of the $\rho$ is considered.
The molecular nature of the $f_2(1270)$ and related tensor states from Ref.~\cite{Geng:2008gx} has been tested in a large number of processes; for instance, in the two-photon decay of the $f_2(1270)$ \cite{Nagahiro:2008um}; 
the two-photon and one-photon-one-vector decays of the $f_2(1270)$, $f'_2(1525)$, and $K^*_2 (1430)$ \cite{Branz:2009cv}; 
the $J/\psi \to \phi (\omega) f_2(1270)$,  $\phi (\omega) f'_2(1525)$ and $J/\psi \to K^*_0(892) \bar K^{*0}_2(1430)$ decays \cite{MartinezTorres:2009uk}; 
the radiative decay of $J/\psi$ into $f_2(1270)$ and $f'_2(1525)$ \cite{Geng:2010kma}; 
the $\psi(2S)$ decays into $\omega (\phi) f_2(1270), \omega (\phi) f'_2(1525)$, $K^*_0(892) \bar K^*_2(1430)$ and the radiative decays of $\Upsilon(1S)$, $\Upsilon(2S)$, $\psi(2S)$ into $\gamma f_2(1270)$, $\gamma f'_2(1525)$, $\gamma f_0(1370)$, and $\gamma f_0(1710)$ \cite{Dai:2013uua,Dai:2015cwa}; 
and the ratio of the decay widths of $\bar B^0_s \to J/\psi f_2(1270)$ and $\bar B^0_s \to J/\psi f'_2(1525)$ \cite{Xie:2014gla}. 
The agreement with experimental data is quite good in general, providing a consistency test for the largely molecular nature of these states.   
Similarly, in Ref.~\cite{Xie:2014twa}, taking the molecular picture for the $f_2(1270)$ resonance, the $\gamma p \to p f_2(1270)$ reaction has been studied and the theoretical results of the differential cross sections are in agreement with the experimental data of Ref.~\cite{CLAS:2009ngd}, providing also support for the molecular picture of the $f_2(1270)$ state in a baryonic reaction.

It is interesting to see the effect of the $K^{*+} K^{*+}$ repulsion.  
To see it, we kill the amplitude $t_2$, for the $K^{*+} K^{*+}$ interaction. 
The results obtained now are shown in Fig.~\ref{Fig:fig2}. 
The peak of the state appears now at $3603\mev$. 
The effect of the $K^{*+} K^{*+}$ repulsion has been to reduce the binding by about $23 \mev$. 
As we can see, the effect of the $K^* D^*$ attraction is much more important than the $K^{*} K^{*}$ repulsion, but the latter has undoubtedly some effect in the precise value of the binding.
\begin{figure}[t]
\begin{center}
\includegraphics[width=0.48\textwidth]{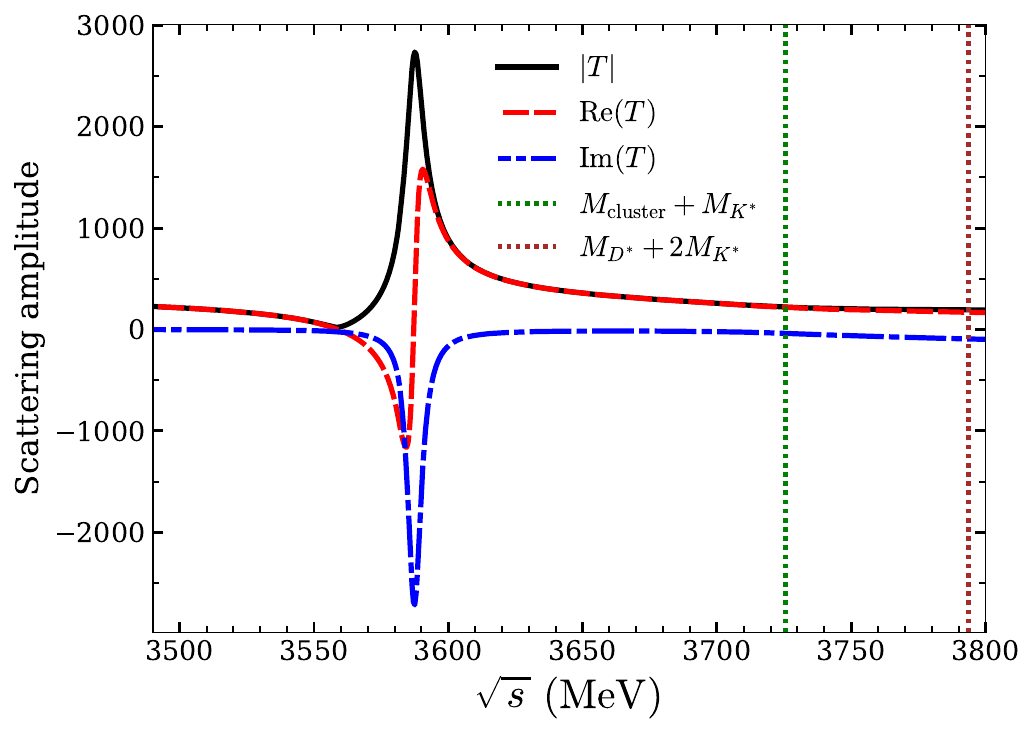}
\end{center}
\vspace{-0.5cm}
\caption{Same as Fig.~\ref{Fig:fig1} but removing the momentum cut implemented through the $\Theta$ functions in Eq.~\eqref{eq:5}, showing the role of this cutoff in shaping the amplitude.}
\label{Fig:fig3} 
\end{figure}

Next we carry another test and see the effect of introducing the $\Theta(q_{\text{max}}^{(i)} - q_i^*)$ functions in Eq.~\eqref{eq:5}. 
The results can be seen in Fig.~\ref{Fig:fig3} where, with respect to Fig.~\ref{Fig:fig1} we have removed the $\Theta(q_{\text{max}}^{(i)} - q_i^*)$ function. 
The peak appears now at $3588 \mev$. 
The consideration of the $\Theta(q_{\text{max}}^{(i)} - q_i^*)$ function in Fig.~\ref{Fig:fig1} decreases the binding energy in $38 \mev$. 
This result is intuitive and one can see it in single channel calculations. 
Indeed, with $T = (V ^{-1}-G)$, with an attractive potential, the reduction of the strength in $G$ as a consequence of the cutoff can be compensated with a reduction of the strength $V^{-1}$, and hence an increase of the strength of $V$, to get the pole at the same position. 
If one keeps the same potential then the binding is smaller.

The tests carried above also tell us that the results are very stable and the binding of about $100 \mev$ with respect to the $K^*$-cluster energy is a solid prediction. 
It would be interesting to see how this state could be observed experimentally.
We can suggest to look at decay modes of the state. 
One of them is to look for the decay of the $K^*$ into $K \pi$, but this would leave another $K^*$ plus $D^*$. 
We find more suitable to look into another decay channel. 
Indeed, in Ref.~\cite{Molina:2022jcd} the decay channels of $K^* D^*$ in $I=1$ and $J=2$ were investigated, and the decay into $K D$ with $L=2$ is possible. 
Actually, in Ref.~\cite{Molina:2022jcd} the width of the $D^*K^*- D^*_s \rho$ state in $I=1$ and $J=2$, was evaluated and was around $19 \mev$, coming both from the decay into $DK$ and the decays of the $\rho$ and the $K^*$, in similar amounts.
This would lead us to  $K D K^*$ in the final state which involves fewer particles to be detected,
and still would have a sizeable decay rate.
The measurement of invariant mass distributions of three particles is a routine work in the LHCb collaboration to identify particles \cite{LHCb:2015yax,LHCb:2020bls},
including vector mesons, like the $\Lambda_b^0 \to \Lambda_c^+ D_s^- \phi$ decay, studied in Ref.~\cite{LHCb:2025lwm}.

One could also think about the decay into baryon-antibaryon. 
With the quark configuration $u \bar s u \bar s c \bar d$, the baryon-antibaryon configuration is $\Sigma_c^{++} \,\bar \Xi^+$, or some of their excited states. 
However, the mass of $\Sigma_c^{++} \,\bar \Xi^+$ is $3776 \mev$, which is more than $100 \mev$ higher than the mass of the state found, and hence, it cannot decay to this channel. 
Another possible decay mode, having fewer particles to detect, would be the $DKK$ mode. 
However, this requires more interaction steps than the production of the $DKK^*$ mode, and should have a smaller production rate. 
The $DK K^*$ decay mode stems as the most likely mode for the detection.

The former discussion concluded that the optimal decay mode is $DKK^*$, but now we would like to discuss how likely is to see in some reaction the peak that we obtain. 
First, let us remember that we obtained a width of the state of the order of $10 \mev$, coming exclusively from the $K^*$ and $\rho$ widths. 
We did not include the width of the $I=1, J=2$ state for decay into $DK$, which was evaluated in Ref.~\cite{Molina:2022jcd}. 
The total width of this state was $19 \mev$, coming from equal amounts from the $K^*$ and $\rho$ decays and the decay of the state into $DK$. 
Hence, we should add about $10 \mev$ from $DKK^*$ decay of the three body system to the $10 \mev$ that we already have from $K^*$ and $\rho$ decays. Hence, we expect about half of the total decay of our three body system into $DKK^*$.

Next we would like to discuss the chances to see that in an experiment. 

We have in mind experiments like in LHCb or ALICE using $pp$ collisions, for instance. 
As mentioned before, the reconstruction of particles through their three or four body decays is routine in LHCb \cite{LHCb:2020bls,LHCb:2015yax,LHCb:2025lwm}. 
It is also becoming available in the study of correlation functions in the ALICE collaboration \cite{DelGrande:2021mju,ALICE:2022boj,ALICE:2023gxp,ALICE:2023bny,Garrido:2024pwi,Garrido:2025lar}. 
Examples in LHCb, in addition to those mentioned above, are $\Lambda_b^0 \to \Lambda_c^+ D_s^- K^+ K^-$ \cite{LHCb:2025lwm} or $B^+_c \to D^+ K^+ \pi^-$ \cite{LHCb:2025qcs}, where in the identification of the $\Lambda_b$ or $B_c$ through these decays one sees a peak with a strength about $4$ times bigger than the background from these final particles coming from other sources than the $\Lambda_b$ or $B_c$ decays. 
One can say that in the production of these particles what one is measuring is the probability that the $\Lambda_b$ or $B_c$ are produced in the $pp$ collisions, because once produced they will decay, yet the branching ratios are of the order of $10^{-4}$ in the first reaction and $10^{-3}$ in the second one, which indicates that the background of uncorrelated final particles in those decays is really small. 
In our case we propose to measure three particles $KDK^*$ which would be coming from the formation of the predicted three particle bound state and its posterior decay, and we might think that this probability is smaller than that of producing a single particle like the $\Lambda_b$ or $B_c$. 
The measure of this probability is provided by the ALICE measurements of the correlation functions of three particles, where distinct correlation function different from unity are seen, for instance in the $ppp$ or $pp\Lambda$ correlation functions measured in Ref.~\cite{ALICE:2022boj}. 
This is also the case in the $p f_1(1285)$ correlation function, where the probability of finding $p K \bar K \pi$ particles is measured.  
In this case, we can take advantage of the results of Ref.~\cite{Encarnacion:2025lyf}, where the three body scattering matrix above threshold, where the correlation functions are measured, is smaller that at the peak below threshold. 
Since the probability  to find these particles is proportional to their wave function squared, or similarly, the $t$ matrix squared, one can conclude that if clear signals are seen for these three body states above threshold, it should be easier to see them below threshold where the peak is predicted. 
Note that this is also the situation in the present case, as one can see in Fig.~\ref{Fig:fig3}, where the strength of the peak below threshold is much bigger than the $t$ matrix around threshold. 
Below threshold one would measure the three particle  bound state through their decay products, in particular the $KDK^*$ mode would account for about half of the strength. 
All these arguments imply that the observation of the predicted peak is already at reach, which should stimulate experimental searches.

\section{Conclusions}
We have investigated the possibility of having a bound state of $K^{*+} D^{*+} K^{*+}$, which is a mesonic state extremely exotic since it has charge three, isospin $3/2$, a quark component of $c\bar d \bar s u \bar s u$, and through flavor conservation in strong interactions it cannot decay into a system of fewer mesons. 
In addition the state has a large angular momentum $J=3$, from the alignment of the spins of the three vectors, which also involve large angular momenta in decays channels. 
All these factors make this state rather stable. 
We looked at the structure of the state by starting from a two body cluster of $D^* K^*$ in $I=1$ and $J=2$, which was found to have a bound state of about $68 \mev$ binding in Ref.~\cite{Molina:2022jcd} (also found bound of Ref.~\cite{Duan:2023lcj}), and then have added another $K^*$ with spin aligned in the same direction as in the original $D^* K^*$, such as to make a $J=3$ state. 
All pairs now have $J=2$, and the external $K^*$ with the $D^*$ of the cluster have again an attractive interaction.
The $K^* K^*$ interaction in $I=1$ and $J=2$ is, however, repulsive, but we find that this repulsion is not enough to overcome the strong $K^* D^*$ attraction in this channel. 

The calculational scheme used here relies upon the fixed center approximation for the interaction of an external particle with a bound cluster, which has been modified to make it unitary at the threshold of the system. 
The new formalism is then closer to the standard technique used in nuclear physics to deal with the interaction of particles with nuclei by means of an optical potential. 
We construct the scattering matrix of an external $K^*$ with the cluster of $K^* D^*$ and find a resonant like state with a typical amplitude structure where the imaginary part has a peak and the real part changes sign around the peak position.
The binding of the state is around $100 \mev$ with respect the mass of a $K^*$ and the cluster of bound $K^* D^*$, and the width obtained is of the order of $10 \mev$, which makes its identification easy in principle. 
In view of the possible decay mode of the cluster of $K^* D^*$ into $KD$, we suggest the measurement of the invariant mass of $K D K^*$, something that is feasible with present techniques in the LHCb collaboration and ALICE among others.

\section*{Acknowledgments}
We thank Dr. Hai-Peng Li for checking some results of this work. 
This work is partly supported by the National Natural Science Foundation of China (NSFC) under Grants No. 12575081 and No. 12365019,
and by the Natural Science Foundation of Guangxi province under Grant No. 2023JJA110076,
and by the Central Government Guidance Funds for Local Scientific and Technological Development, China (No. Guike ZY22096024).
This work was supported by the National Key R{\&}D Program of China (Grant No. 2024YFE0105200).
R. M. acknowledges support from the ESGENT program with Ref. ESGENT/018/2024 and the PROMETEU program with Ref. CIPROM/2023/59, of the Generalitat Valenciana, and also from the Spanish Ministerio de Economia y Competitividad and European Union (NextGenerationEU/PRTR) by the grants with Ref. CNS2022-136146 and PID2023-147458NB-C21, funded by MICIU/AEI/10.13039/501100011033. 
This work is also partly supported by the Spanish Ministerio de Economia y Competitividad (MINECO) and European FEDER funds under Contracts No. FIS2017-84038-C2-1-PB, PID2020-112777GB-I00, and by Generalitat Valenciana under contract PROMETEO/2020/023. 
This project has received funding from the European Union Horizon 2020 research and innovation program under the program H2020-INFRAIA-2018-1, grant agreement No. 824093 of the STRONG-2020 project.

\appendix
\section{$D^{*} K^{*}$ and $K^{*}K^{*}$ amplitudes}
\subsection{$D^{*} K^{*}$ amplitude in $I=1$ and $J=2$}

In the paper~\cite{Molina:2010tx}  we have the transition potentials, $V_{ij}$, between the two single channels $D^{*}K^{*}$ and $D_{s}^{*}\rho$ in $I=1$ and $I=2$ given in Table \ref{Tab:p4} below,
\begin{table*}[htbp]
\centering
\caption{Amplitudes for $C=1$, $S=1$ ,$I=1$ and $J=2$ \cite{Molina:2010tx}.}
\label{Tab:p4}
\setlength{\tabcolsep}{10pt}
\begin{tabular}{ccc}
\hline\hline
& $K^*D^*$ & $D_s^*\rho$ \\[2mm]
\hline
$K^*D^*$ & $\dfrac{g^2(p_1+p_3)\,(p_2+p_4)}{2}\;(\dfrac{1}{m_{\rho}^2}-\dfrac{1}{m_{\omega}^2 })$ & $-2g^2-\dfrac{g^2\,\left( p_1+p_4 \right)\,\left( p_2+p_3 \right)}{m_{D^{*}}^{2} }-\dfrac{g^2\,\left( p_1+p_3 \right)\,\left( p_2+p_4 \right)}{m_{K^{*}}^{2}}$ \\[5mm]
$D_s^*\rho$ &  & $0$\\
\hline\hline
\end{tabular}
\end{table*}
where $g = M_{V} / 2 f$ ($M_{V} = 800 \mev$, $f = 93 \mev$) and $p_1, p_2$ are the $D^{*}, K^{*}\,(D_{s}^{*},\rho)$ initial momenta and $p_3, p_4$ the final $D^{*},K^{*} \,(D_{s}^{*},\rho)$ momenta. Projected over $S$-wave one has
\begin{equation}
	\begin{split}
		(p_{1} + p_{3}) \cdot (p_{2} + p_{4}) \to& \frac{3}{2}s - \frac{1}{2}(m_{1}^{2} + m_{2}^{2} + m_{3}^{2} + m_{4}^{2}) \\
		&\!\!\!\!-\frac{1}{2s}(m_{1}^{2} - m_{2}^{2})(m_{3}^{2} - m_{4}^{2}),~~~
	\end{split}
\end{equation}
\begin{equation}
	\begin{split}
		(p_{1} + p_{4}) \cdot (p_{2} + p_{3}) \to& \frac{3}{2}s - \frac{1}{2}(m_{1}^{2} + m_{2}^{2} + m_{3}^{2} + m_{4}^{2})\\
		&\!\!\!\!+ \frac{1}{2s}(m_{1}^{2} - m_{2}^{2})(m_{3}^{2} - m_{4}^{2}).
	\end{split}
\end{equation}
Then in the coupled channels matrix form we have the scattering matrix
\begin{equation}
	T=[1-VG]^{-1}\, V,
\end{equation}
and $G$=diagonal$\big[G_{1},G_{2}\big]$ (with the index $i=1, 2$ standing for $D^{*}K^{*}$ and $D_{s}^{*}\rho$ channels respectively), with 
\begin{eqnarray}\label{eq:Gcut1}
	G_i(s) = \int_{|{\vec q\,}| < q_{\rm max}} \,&&\dfrac{\dd^3 q}{(2\pi)^3}  \;\dfrac{\omega^{(i)}_1(\vec q\,) + \omega^{(i)}_2(\vec q\,)}{2 \,\omega^{(i)}_1(\vec q\, ) \; \omega^{(i)}_2(\vec q\,)} \nonumber\\[2mm]
		&&\!\!\!\!\!\!\!\!\!\!\times\dfrac{1}{s-\left[\omega^{(i)}_1(\vec q\,) + \omega^{(i)}_2(\vec q\,)\right]^2+i\epsilon},~~
\end{eqnarray}
where $\omega^{(i)}_j(\vec q\,) = \sqrt{{\vec{q}}^{\;2}+{m^{(i)}_{j}}^{2}}$, with $m^{(i)}_{1},m^{(i)}_{2}$ the masses of the particles in the intermediate states of the loop of the $i$ channel. However, in order to take into account the $\rho$ and $K^{*}$ widths we do a convolution of these $G$ functions with the spectral functions of the $
\rho$ or $K^{*}$,
\begin{eqnarray}\label{eq:A5}
	G_{i}(s)\to\tilde{G}_{i}(s)&=&\frac{1}{N}_{i}\int\limits^{(m_i+3.5\Gamma_i)^2}\limits_{(m_i-3.5\Gamma_i)^2}
	d\tilde{m}^2_i\left(-\frac{1}{\pi}\right) \nonumber\\
	&&\!\!\times\mathrm{Im}\frac{1}{\tilde{m}_{i}^2-m_i^2+i\Gamma_i\tilde{m}_i}\cdot G_{i}(s,\tilde{m}_i),~~~~~~
\end{eqnarray}
where
\begin{eqnarray}
	N_i=\int\limits^{(m_i+3.5\Gamma_i)^2}\limits_{(m_i-3.5\Gamma_i)^2}
	d\tilde{m}^2_i\left(-\frac{1}{\pi}\right)\mathrm{Im}\frac{1}{\tilde{m}_{i}^2-m_i^2+i\Gamma_i\tilde{m}_i},~~~~~
\end{eqnarray}
with $m_{1}= m_{K^{*}}, \Gamma_1=\Gamma_{K^{*}}$ for the $K^*D^*$ channel, and $m_{2}=m_{\rho}, \Gamma_2=\Gamma_\rho$ for the $D^*_s \rho$ channel.

For the FCA calculation we need the $T_{D^{*}K^{*},D^{*}K^{*}}$ amplitude that we called $t_1$ in the text.

\subsection{$K^* K^*$ amplitude in $I=1$ and $J=2$}

\begin{figure}[b]
\begin{center}
\includegraphics[width=0.48\textwidth]{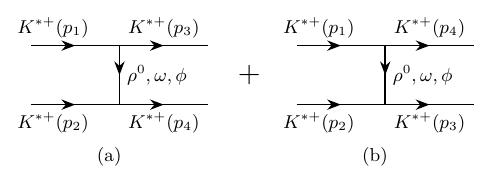}
\end{center}
\vspace{-0.7cm}
\caption{Diagrams included in the calculation of the $K^{*+}K^{*+}\to K^{*+}K^{*+}$ potential.}
\label{Fig:fig22}
\end{figure}

This amplitude is not available in the literature, and we have calculated it here using the local hidden gauge approach~\cite{Bando:1984ej,Bando:1987br,Meissner:1987ge,Nagahiro:2008cv}.
The amplitude is calculated from the diagrams of Fig.~\ref{Fig:fig22}.

The diagram for the $VVV$ vertex in the diagrams of Fig.~\ref{Fig:fig22} is given in the local hidden gauge approach for small moments of the external particles as
\begin{eqnarray}
	\mathcal{L}= ig \langle( V_{\mu } \partial_{\nu} V^{\mu} - \partial_{\nu} V^{\mu} V_{\mu} )V^{\nu}\rangle.
\end{eqnarray}
The vector meson matrix $V$ is
\begin{equation}
	\renewcommand{\tabcolsep}{1cm}
	\renewcommand{\arraystretch}{2}
	V=\left(
	\begin{array}{ccc}
		\frac{\omega+\rho^0}{\sqrt{2}} & \rho^+ & K^{*+}\\
		\rho^- &\frac{\omega-\rho^0}{\sqrt{2}} & K^{*0}\\
		K^{*-} & \bar{K}^{*0} &\phi
	\end{array}
	\right).
\end{equation}
Note that in the Lagrangian, $V^{\nu}$ has to be the exchanged vector. Indeed, if it were an external vector, then the $\nu$ indices would be spatial, $\nu=1,2,3$, since $\epsilon ^0=0$ for a particle at rest, that we have assumed for the external vectors. 
Then $\partial_{\nu}$ will give a three momentum, but that is zero from the starting assumptions. Then it is easy to see that we obtain the transition potential for $K^{*+}K^{*+} \to K^{*+}K^{*+}$
\begin{equation}
	\begin{aligned}
		V_{ex} =& 2\, \dfrac {g^2} {M_V^2}\big[(p_1 + p_3)\cdot (p_2 + p_4)\,\epsilon_{\nu}(1) \, \epsilon_{\alpha}(2) \,\epsilon^{\nu}(3) \,  \epsilon^{\alpha}(4)\\
		&\!\!+ (p_1 + p_4)\cdot (p_2 + p_3) \;\epsilon_{\nu}(1) \,  \epsilon_{\alpha}(2) \, \epsilon^{\alpha}(3) \, \epsilon^{\nu}(4)\big],
	\end{aligned}
\end{equation}
and recalling the spin projectors from Ref.~\cite{Molina:2008jw} in terms of the four polarization vectors,
\begin{eqnarray}\label{eq:projector}
	\mathcal{P}^{(0)}&=&\frac{1}{3}\epsilon_\nu(1)\,\epsilon^\nu(2)\,\epsilon_\alpha(3)\, \epsilon^\alpha(4),\nonumber\\
	\mathcal{P}^{(1)}&=&\frac{1}{2}\big[\epsilon_\nu(1)\,\epsilon_\alpha(2)\,\epsilon^\nu(3)\,\epsilon^\alpha(4)-
	\epsilon_\nu(1)\,\epsilon_\alpha(2)\,\epsilon^\alpha(3)\,\epsilon^\nu(4)\big],\nonumber\\
	\mathcal{P}^{(2)}&=&\frac{1}{2}\big[\epsilon_\nu(1)\,\epsilon_\alpha(2)\,\epsilon^\nu(3)\,\epsilon^\alpha(4)+
	\epsilon_\nu(1)\,\epsilon_\alpha(2)\,\epsilon^\alpha(3)\,\epsilon^\nu(4)\big]\nonumber\\
	&&-\frac{1}{3}\epsilon_\nu(1)\,\epsilon^\nu(2)\,
	\epsilon_\alpha(3)\,\epsilon^\alpha(4).
\end{eqnarray}
We easily obtain
\begin{eqnarray}
	V_{ex}^{(J=2)}&=& 2\, \dfrac{g^2}{M_V^2}\big[(p_1 + p_3)\cdot (p_2 + p_4)+(p_1 + p_4)\cdot (p_2 + p_3) \big]\nonumber\\
	&=& 2\, \dfrac {g^2} {M_V^2}\;[3s-\left( m_1^2+m_2^2+m_3^2+m_4^2\right)].
\end{eqnarray}
The local hidden gauge approach has also a contact term
\begin{equation}
	\mathcal{L}_\mathrm{VVV}=\frac{1}{2}g^2\langle [V_\mu,V_\nu]V^\mu V^\nu\rangle,
\end{equation}
which, again, projecting over $J=2$ gives the transition potential
\begin{equation}
	V_{c}^{(J=2)}=4g^{2}.
\end{equation}
As one can see, both the $V_{ex}^{(J=2)}$ and $V_{c}^{(J=2)}$ potentials are repulsive. 
We construct the $T$ matrix for the $K^*K^*$ single channel as
\begin{equation}\label{eq:T}
	T=\dfrac{V}{1-V\,\frac{1}{2}\,G},
\end{equation}
with $V=V_{ex}^{(J=2)}+V_{c}^{(J=2)}$, and the $G$ function as in  Eq.~\eqref{eq:Gcut1} for $K^{*}K^{*}$ propagation.
The factor $\frac{1}{2}$ in $\frac{1}{2}\,G$ of Eq.~\eqref{eq:T} is due to the identity of the two $K^{*+}$ particles.
We use the same $q_\mathrm{max}$ as in the $D^{*}K^{*}$ case, but $VG$ being negative the denominator in Eq.~\eqref{eq:T}, $1-\frac{1}{2}VG$ only produces moderate changes in $V$.
In addition, the arguments $s_1, s_2$ of the $ t_{1} $ and $ t_{2} $ amplitudes are evaluated in Refs.~\cite{Encarnacion:2025lyf,Ikeno:2025bsx,Agatao:2025ckp} and are given by
\begin{equation}\label{eq:A15}
	\begin{split}
	s_{1}(K^{*}D^{*}) = m_{K^{*}}^{2} + (\xi \, m_{D^{*}})^{2} + 2\,\xi \,m_{D^{*}}q^{0},\\[2.5mm]
	s_{2}(K^{*}K^{*}) = m_{K^{*}}^{2} + (\xi \,m_{K^{*}})^{2} + 2\,\xi \, m_{K^{*}}q^{0},
	\end{split}
\end{equation}
with $q^{0}=\frac{s-m_{K^{*}}^{2}-M_{c}^{2}}{2M_{c}}$; $\xi=\frac{M_{c}}{m_{D^{*}}+m_{K^{*}}},$ where $q^{0}$ is the external $K^{*}$ energy in the cluster rest frame and $\xi$ is a parameter introduced to account for the binding of the cluster, splitting to binding between the particles of the cluster proportionally to their mass.

\bibliographystyle{a}
\bibliography{refs}
\end{document}